\documentstyle[12pt]{article}
\begin{document}
\setlength{\baselineskip}{26pt}

\begin{titlepage}
\centerline{\LARGE Corner Transfer Matrix }
\centerline{\LARGE Renormalization Group Method }

\author{T. Nishino}
\vskip 15pt
\centerline{ T. Nishino$^{1}$ and K. Okunishi$^{2}$}
\centerline{\sl $1$ Physics Department, Graduate School of Science, }
\centerline{\sl Tohoku University, Sendai 980-77, Japan }
\centerline{\sl $2$ Physics Department, Graduate  School of Science, }
\centerline{\sl Osaka University,  Toyonaka, Osaka 560, Japan }
\vskip 20pt

\begin{abstract}
\setlength{\baselineskip}{24pt}
We propose a new fast numerical renormalization group method,
the corner transfer matrix renormalization group (CTMRG) method,
which is based on a unified scheme of Baxter's corner  transfer
matrix method and White's density matrix renormalization group
method. The key point  is that a product of four corner transfer
matrices coincides with the density  matrix. We formulate the CTMRG
method
as a renormalization of 2D classical models.
\end{abstract}

\begin{itemize}
\item[\bf PACS codes:]{05.50.+q, 02.70.-c, 75.10.Hk.}
\end{itemize}
\end{titlepage}
\newpage


The renormalization group is one of the basic concepts in physics
\cite{Kadanoff,Wilson}. Real space representation of the
renormalization
group --- the real space renormalization group --- has been
applied to various lattice models \cite{Burkhardt}. Recently,
White established a numerical renormalization algorithm, which is
referred as `density matrix renormalization group (DMRG) method'
\cite{White1,White2}. The method has been applied to various
one-dimensional (1D) quantum lattice models
\cite{White2,Yu,Noack}, because it is possible to treat large scale
systems with relatively small numerical calculation.

Although DMRG method was originally proposed as a renormalization
procedure for 1D quantum systems, the method has an implicit relation
with 2D classical models. \"Ostlund and Rommer analyzed the
thermodynamic
limit of the DMRG method  \cite{Ostlund}, and pointed out that the
method
is a mapping from 1D quantum lattice models to effective classical
lattice models. They show that the orthogonal matrix, which represents
the
block-spin transformation, plays a role of a transfer matrix of the
classical model. What is the classical model, then? Roughly speaking,
the
model corresponds to a 2D square-lattice  model, which is obtained
through the Trotter decomposition \cite{Trotter,Suzuki} of  the
operator
$exp(-\beta {\hat H})$ \cite{density}; the row-to-row transfer matrix
corresponds to the imaginary time shift operator $exp(-\Delta\beta
{\hat
H})$; the transfer matrix discussed by \"Ostlund and Rommer is a
renormalized column-to-column transfer matrix.

The relation between the DMRG method and 2D classical systems leads a
new
view point. {\it We find that the density matrix is expressed as a
product of Baxter's corner transfer matrices} (CTMs)
\cite{Baxter1,Baxter2,BaxterT}. Moreover, the DMRG method and Baxter's

variational method on CTM have many aspects in common; both of
them is a natural extension of the Kramers-Wannier approximation
\cite{Kramers}. From this unified view point, we present a very fast
numerical renormalization procedure for 2D classical systems: the
corner
transfer matrix renormalization group (CTMRG) method. At first, we
briefly review the way to apply the DMRG method to 2D classical
systems
\cite{Nishino} in order to see the advantage of the CTMRG method. We
choose the `interaction round a face (IRF)  model' as an example of 2D
classical models. We then present the theoretical background and the
numerical algorithm of the CTMRG method. The numerical superiority of
the CTMRG method is demonstrated by a trial calculation on the Ising
model. We finally discuss the way to apply the CTMRG method to 1D
quantum
systems.

The IRF model includes various 2D lattice models such as
the Ising model and the Potts model \cite{BaxterT}. The IRF model is
defined by the Boltzmann weight $W(a'b'|ab)$ on each face --- a square
surrounded by four $n$-state spins $a$, $a'$, $b$, and $b'$. The
row-to-row transfer matrix of the IRF model is expressed as
\begin{equation}
T(a'b'c' \ldots y'z'|abc \ldots yz) =
W(a'b'|ab) W(b'c'|bc) \ldots W(y'z'|yz) ,
\end{equation}
where the position of the spin variables are shown in Fig.1.
Throughout
this paper, we assume that $W$ is isotropic and symmetric ---
$W(ab|cd)
= W(ba|dc) = W(ca|db) = W(dc|ba)$ --- in order to simplify the
discussion. Generalizations to anisotropic, and/or asymmetric cases
are
straightforward.

The DMRG method for the IRF model is expressed as a renormalization of
the
transfer matrix
\begin{equation}
T(a'b'c' \ldots y'z'|abc \ldots yz) \rightarrow
P(i'\xi'|i\xi) P(i'\eta'|i\eta)
\end{equation}
as shown in Fig.1, where $P$ represents a renormalized transfer matrix
for
the left/right-half 2D lattice. Hear after, we call $P$ `partial
transfer
matrix'. The greek indices $\xi,\xi',\eta$ and $\eta'$ are $m$-state
block-spin variables, that are shown by open  squares. The
renormalized
transfer matrix for the lattice with two additional columns is $T' =
P\cdot W\cdot W\cdot P$, where the dot `$\cdot$' denotes a scalar
product. The eigenvalue equation for $T'$ is
\begin{equation}
\sum_{\xi i j k \eta} P(i'\xi'|i\xi) W(i'j'|ij) W(j'k'|jk)
P(k'\eta'|k\eta)  V(\xi ijk\eta) = \lambda\, V(\xi'i'j'k'\eta') ,
\end{equation}
where $\lambda$ is the non-degenerate largest eigenvalue of $T'$, and
$V$
is the corresponding eigenvector. What is called  `density
matrix' $\rho$ is defined by a product \cite{eigenv}
\begin{equation}
\rho(\xi'i'|\xi i) = \sum_{jk\eta} V(\xi'i'jk\eta) V(\xi ijk\eta) ,
\end{equation}
where we have used the fact that $V$ is real. The DMRG method is a
systematic procedure to obtain $P$ by using the information of $\rho$.

At this stage, we explain our physical view of the density matrix.
Since
$\lambda$ in Eq.3 is the largest eigenvalue, $V$ is given by the large
$L$
limit of $(T')^L X$, where $X$ is a vector that is not orthogonal to
$V$.
Therefore, the vector $V(\xi ijk\eta)$ represent the Boltzmann weight
for
the lower (or the upper) half 2D lattice with the spin configuration
$\{\xi ijk\eta\}$ on the horizontal boundary. Equation 4 indicates
that
$\rho$ is created by partially joining the two halves of the 2D
lattice,
as shown in Fig.2; The density matrix $\rho$ represents the entire
system
with a cut.

The physical background of the density matrix enables us to skip
the eigenvalue problem (Eq.3). {\it What is really necessary in order
to
obtain $\rho$ is not the eigenvector of $T'$, but is the Boltzmann
weight
that stands for the upper/lower half lattice.} How can we get $\rho$,
then? We employ Baxter's corner transfer matrix (CTM)
\cite{Baxter1,Baxter2,BaxterT} for this purpose. He expresses the
half-infinite lattice as a product of CTMs
\begin{equation}
V(\xi ijk\eta) \approx \sum_{l\beta}
A'(jk\eta|jl\beta) A'(jl\beta|ji\xi) ,
\end{equation}
where $A'(i'j'\alpha'|ij\alpha)$ is the CTM that represents the
Boltzmann
weight for a quadrant (or corner) of the 2D lattice. (See Fig.3.) The
element $A'(i'j'\alpha'|ij\alpha)$ is zero when $i' \ne i$.  The
notation
`$\approx$' denotes that the r.h.s. of Eq.5 is not the same as the
eigenvector in Eq.3, but is approximately exact. We further decompose
$V$
into a fine product form, $V \approx (P\cdot W\cdot W\cdot P)(A\cdot
P\cdot P\cdot A)$, as shown in Fig.3. The relation between $A$ and
$A'$
is
\begin{equation}
A'(jk\eta|jl\beta) = \sum_{m\mu\alpha}
W(jk|lm) P(k\eta|m\mu) P(l\beta|m\alpha) A(m\alpha|m\mu) ,
\end{equation}
where we have used the symmetry of the Boltzmann weight $W$. The
factor
$W\cdot P\cdot P$ in Eq.6 is a kind of transfer matrix that acts on
$A$,
and increases the size of the corner. Substituting Eq.5 into Eq.4, we
get
a new expression of $\rho$
\begin{equation}
\rho(\xi'i'|\xi i) \approx \sum_{jklm\alpha\beta\gamma}
A(ji'\xi'|jk\alpha) A(jk\alpha|jl\beta)
A(jl\beta|jm\gamma) A(jm\gamma|ji\xi) .
\end{equation}
The density matrix is a product of four CTMs. The relation
between $\rho$ and $A$ in Eq.7 unifies Baxter's  CTM method and
White's
DMRG method.

Now we explain the key point of our new numerical method, which is a
self-consistent relation between $A$ and $A'$. The relation consists
of
the mapping from $A$ to $A'$ (Eq.6) and the renormalization from $A'$
to $A$
\begin{equation}
\sum_{jj'\eta\eta'} O^{T}(\xi'|j'\eta') A'(i'j'\eta'|ij\eta)
O(j\eta|\xi)
\rightarrow A(i'\xi'|i\xi)
\end{equation}
together with the renormalization of $P$
\begin{equation}
\sum_{jj'\eta\eta'} O^{T}(\xi'|j'\eta') W(i'j'|ij)
P(j'\eta'|j\eta) O(j\eta|\xi) \rightarrow P(i'\xi'|i\xi) .
\end{equation}
The orthogonal matrix $O$ represents the block spin
transformation, and is obtained from the diagonalization of
$\rho$ \cite{singular}
\begin{equation}
\sum_{ii'\xi\xi'} O^{T}(\eta|i'\xi') \rho(i'\xi'|i\xi) O(i\xi|\zeta)
= \delta_{\eta\zeta} \omega_{\eta} ,
\end{equation}
where $O^{T}$ is the matrix transpose of $O$, and
$\omega_{\eta}$ is the $m$-numbers of eigenvalues of $\rho$ from the
largest \cite{White1,White2}. The self-consistent relation for CTM
(Eq.6-10) has the same solution as Baxter's CTM method
\cite{Baxter1,Baxter2,BaxterT}. We solve the self-consistent relation
by
way of the following numerical procedure: (I) Set appropriate initial
values for $P$ and $A$ according to the boundary conditions: (II)
Obtain
$A'$ by using Eq.6: (III) Create  $\rho$ by Eq.7: (IV) Diagonalize
$\rho$
and obtain $O$ (Eq.10): (V) Renormalize $P$ and $A$ according to
Eq.8 and Eq.9: (VI) Repeat (II)-(V) until $A$ and $P$ reach their
fixed
point. We call the method `corner transfer matrix renormalization
group (CTMRG) method', since the renormalization is done for CTM.
After we
obtain $P$ and $A$ at the fixed point, we calculate spin correlation
functions. For example, two-point spin correlation functions along a
row
(or column) is obtained from several large eigenvalues of $P \cdot P$
in the r.h.s. of Eq.2. It is also possible to calculate correlation
functions by using the fixed point value of $O$ in Eq.10
\cite{Ostlund}.

We apply the CTMRG method to square lattice Ising model, which is a
special case of the IRF model. Figure 4 shows the calculated local
energy
$E(T)$ --- the nearest-neighbor spin correlation function --- when $m
=
98$. The data shown by the black dots are obtained after $10 \sim
1000$
iterations, where the data deviate from the exact ones \cite{Onsager}
at
most $10^{-7}$. The numerical precision can be improved by additional
iterations. At the critical temperature, we estimate $E(T_c)$ by
observing its convergence with respect to $N$, which is the number of
iteration. (See inset of Fig.4.) A simple $1/N$ fitting gives $E(T_c)
=
0.70704$, which is close to the exact one $1 / \sqrt{2} = 0.70711$. It
should be noted that the CTMRG method is much faster than the DMRG
method. The CTMRG method requires 8.7sec to obtain $E(T)$ at $T = 2.2$
by
the NEC SX-3 super computer, while the DMRG method consumes 149.8sec
to
deduce the data of the comparable numerical precision \cite{Nishino}.
The
CTMRG method runs faster because it crates $\rho$ by using Eq.7, which
consists of a few $n^2m$-dimensional matrix multiplication. On the
contrary, the DMRG method requires the solution of the
$n^2m^2$-dimensional eigenvalue problem.

We have presented the CTMRG method as a fast numerical renormalization
group method for 2D classical models. We finally discuss a way to
apply
the CTMRG method to 1D quantum lattice models. A natural extension is
given through a mapping from 1D quantum models to 2D classical ones
via
the Trotter formula \cite{Trotter,Suzuki}. The corresponding 2D model
is
a checkerboard type one, which is an anisotropic IRF model. Therefore,
it
is possible to apply the CTMRG method to the 1D quantum models that
have
been analyzed by QMC simulations. It should be noted that the CTMRG
method is free from the sign problem that occasionally makes the QMC
simulation difficult. In principle, correlation functions for both
space
and imaginary time directions can be calculated, since the formulation
of
the CTMRG method is symmetric for both space and imaginary time
directions. We have obtain another extension of the CTMRG method for
1D
quantum systems. This one does not require the Trotter formula, and is
written in a renomalization group on wave functions. The detail will
be
discussed elsewhere \cite{TO}.

The authors would like to express their sincere thanks to Y.~Akutsu
and
M.~Kikuchi for valuable discussions and encouragement. T.~N. thank to
S.~R.~White for helpful comments and discussions about applications of
the DMRG method to 2D lattice models. Numerical calculations were done
by
NEC SX-3/14R in computer center of Osaka University. The present work
is
partially supported by a Grant-in-Aid  from Ministry of  Education,
Science and Culture of Japan.  
\newpage

\begin{figure}
\caption{The row-to-row transfer matrix $T$ of the IRF model is a
product
of the Boltzmann weights on each face. The DMRG  method maps $T$ into
a
product form in Eq.2.}
\end{figure}

\begin{figure}
\caption{The density matrix $\rho$ as a product of $V$. We
regard  $\rho$ as a Boltzmann weight for the entire 2D lattice with a
cut.}
\end{figure}

\begin{figure}
\caption{The Boltzmann weight for the half-infinite lattice is
constructed as a product of corner transfer matrices. The further
decomposition in Eq.6 is also shown.}
\end{figure}

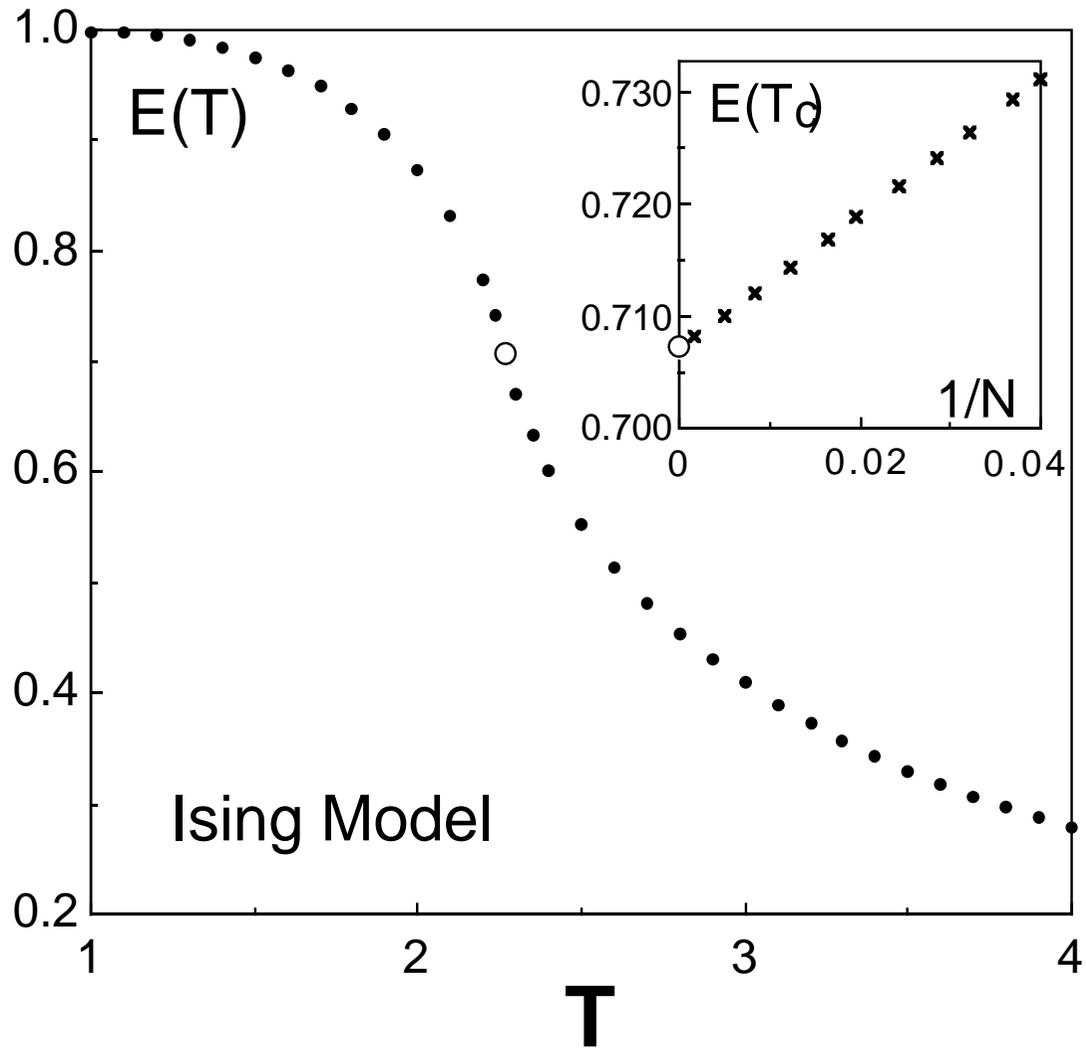
\begin{figure}
\caption{Local energy $E(T)$ of the Ising model. We determine $E(T_c)$
by
observing the convergence with respect to $N$, the number of
iterations
of the CTMRG procedure.}
\end{figure}


\begin{thebibliography}{99}
\bibitem{density} They call the operator $exp(-\beta
{\hat H})$ `density matrix', but in this letter we never do so
in order to avoid a terminological confusion.
\bibitem{eigenv} If the transfer matrix $T'$ is asymmetric, then
$\rho$
is defined by a contraction between the right and the left
eigenvectors
of $T'$.
\bibitem{singular} For the asymmetric IRF model, one has to employ the
singular value decomposition instead of the matrix diagonalization.
\end{thebibliography}
\end{document}